\documentclass[journal]{IEEEtran}
\usepackage{amsmath,amssymb,amsfonts,amsthm}
\usepackage{graphicx}
\usepackage{enumitem}
\usepackage{cite}
\usepackage{multirow}
\usepackage[hidelinks]{hyperref}
\newtheorem{theorem}{Theorem}
\newtheorem{proposition}{Proposition}
\newtheorem{corollary}{Corollary}
\theoremstyle{definition}

\newtheorem{definition}{Definition}

\begin{document}
\title{Rateless Lattice Codes over Fading Wiretap Channels: Finite-Length Secrecy and the Secret-Key Capacity of Degraded Sessions}
\author{Hassan~Khodaiemehr, Khadijeh~Bagheri, and Chen~Feng%
\thanks{The authors are with the School of Engineering, University of British Columbia, Okanagan Campus, Kelowna, BC, Canada (e-mails: hassan.khodaiemehr@ubc.ca, khadijeh.bagheri@ubc.ca, chen.feng@ubc.ca).}}
\maketitle

\begin{abstract}
In rateless transmission the transmitter keeps sending until the legitimate receiver (Bob) can decode. Over a wiretap channel this has a useful side effect: the eavesdropper (Eve) sees only as many symbols as Bob needs, although the transmitter knows neither channel. We study the secrecy of rateless lattice codes in this setting. First, using a public dither and privacy amplification, we prove a finite-length secrecy bound, in total variation distance, for every channel realization, even if Eve knows all codes and seeds and combines all packets. Second, stopping when Bob acknowledges successful decoding, as in hybrid ARQ, costs at most the logarithm of the number of possible feedback outcomes in secret bits. Third, for degraded sessions with state-based stopping and without a pre-shared key or other public discussion, the secret-key rate of a session cannot exceed its accumulated Gaussian secrecy capacity, and adaptive key generation with shaped rateless lattice codes comes within any $\epsilon>0$ of this limit when Eve's channel states are revealed after the session. If only their statistics are known, choosing the key length after the session guarantees a prescribed secrecy-outage probability. Finally, over block Rayleigh fading with a CRC-based acknowledgement and joint belief-propagation decoding, and with Eve $20$~dB weaker on average, the scheme reaches $59$--$69\%$ of the outage-weighted throughput of an ideal infinite-length secure hybrid-ARQ benchmark for Bob SNRs of $8$--$20$~dB.
\end{abstract}

\begin{IEEEkeywords}
Rateless codes, lattice codes, physical-layer security, privacy amplification, secret-key agreement, secure HARQ, block fading.
\end{IEEEkeywords}

\section{Introduction}
\IEEEPARstart{I}{n} rateless transmission, the number of symbols that a passive eavesdropper (Eve) observes is set by the legitimate receiver's (Bob's) channel: when it is good, the transmitter (Alice) stops early, before Eve accumulates much information, without channel knowledge at Alice. Two questions arise. In hybrid ARQ, Bob acknowledges when his decoder succeeds; the stopping time then depends on the message and is itself a side channel; secure hybrid ARQ was analysed in~\cite{TLSP2009} with Gaussian binning in the limit of long blocks, and finite-length limits are known for fixed-length wiretap codes~\cite{YSP2019}, but neither covers a practical rateless code stopped by a decoding-based acknowledgement. And what secret rate can such a scheme achieve per session? Lattice wiretap codes of fixed dimension~\cite{19silva} and fading secrecy capacity with channel knowledge~\cite{BBRM2008,LPS2008} do not answer it.

We study both for rateless lattices~\cite{TITcomp} built from Construction~D$'$ with nested Raptor codes:
\begin{enumerate}[leftmargin=*]
\item a per-state, finite-length secrecy bound that stays valid when Bob acknowledges on decoding success, at a cost of at most $\log_2$ of the number of feedback outcomes (Theorem~\ref{thm:secrecy});
\item the secret-key capacity of degraded sessions: a converse (Theorem~\ref{thm:converse}) and its attainment by adaptive key generation (Corollary~\ref{cor:seccap});
\item a key length with guaranteed secrecy outage (Proposition~\ref{prop:adapt}), and what the spread-spectrum (DSSS) layer of~\cite{arxivPrev} can add (Proposition~\ref{prop:dsss});
\item results with belief-propagation (BP) decoding of dithered transmissions over block Rayleigh fading.
\end{enumerate}

\section{System Model}\label{sec:model}
\emph{Rateless lattices.} Let $\pi_n$ keep the first $n$ coordinates of a vector.
\begin{definition}[Rateless lattice~\cite{TITcomp}]\label{def:rateless}
A sequence $(\Lambda_n)_{n\ge N}$ of full-rank lattices $\Lambda_n\subset\mathbb R^n$ is a \emph{$q$-ary rateless lattice} if (i) $q\mathbb Z^n\subseteq\Lambda_n\subseteq\mathbb Z^n$ for every $n$; (ii) $\pi_n(\Lambda_{n'})=\Lambda_n$ for all $n'\ge n\ge N$; and (iii) the codebook $\mathcal S_n=\Lambda_n\cap\{0,\dots,q-1\}^n$ has a size that does not depend on $n$.
\end{definition}

Integrality makes $\mathcal S_n$ a set of representatives of $\Lambda_n/q\mathbb Z^n$, hence $|\mathcal S_n|=q^n/\mathrm{vol}(\Lambda_n)$, and with (ii)--(iii) it makes $\pi_n:\mathcal S_{n'}\to\mathcal S_n$ bijective~\cite{TITcomp}: a message selects one point of every $\mathcal S_n$, and these are prefixes of one stream. We use the Raptor lattice of~\cite{TITcomp}. Let $\mathcal C_{\mathrm{pre}}\subseteq\mathbb F_2^N$ be a precode of dimension $K$ with parity-check matrix $\mathbf H_{\mathrm{pre}}$, and let $\mathbf s_1,\mathbf s_2,\dots\in\mathbb F_2^N$ be the rows of an LT stage. Then
\begin{equation}\label{eq:raptor}
\begin{aligned}&\Lambda_n=\big\{\mathbf x\in\mathbb Z^n:\ \mathbf H_{\mathrm{pre}}\mathbf x_{[1:N]}\equiv\mathbf 0\!\!\pmod2,\\ &\qquad\ \mathbf s_t^{\mathsf T}\mathbf x_{[1:N]}+x_{N+t}\equiv0\!\!\pmod4\ \text{for}\ t\le n-N\big\}\end{aligned}
\end{equation}
is an integral $4$-ary rateless lattice with $\mathrm{vol}(\Lambda_n)=2^{2n-k}$ and $|\mathcal S_n|=2^k$, $k=K+N$~\cite[Thm.~1]{TITcomp}. A word $\mathbf u=(\mathbf m_0,\mathbf m_1)\in\mathcal C_{\mathrm{pre}}\times\mathbb F_2^N$ is encoded as $\boldsymbol\xi=\mathbf m_0+2\mathbf m_1$ followed by $x_{N+t}=(-\mathbf s_t^{\mathsf T}\boldsymbol\xi)\bmod4$, $t=1,2,\dots$; we denote this stream by $\mathbf x_{\mathbf u}$.

\emph{Channels.} A symbol $a\in\{0,1,2,3\}$ is sent as the centred $4$-PAM amplitude $a-\frac32$, of power $P=\frac54$; constant shifts known to the receivers do not affect distances or information. Bob and Eve observe $Y_{\mathrm B,j}=X_j+\sigma_{\mathrm B,j}Z_j$ and $Y_{\mathrm E,j}=X_j+\sigma_{\mathrm E,j}Z_j'$ with independent standard Gaussian noises, $\sigma=\infty$ for an erasure, and $\mathrm{SNR}_{u,j}=P/\sigma_{u,j}^2$. The states $\mathsf S=((\sigma_{\mathrm B,j}),(\sigma_{\mathrm E,j}))$ have an arbitrary joint law, independent of the message, noises and public randomness, unknown to Alice. We write $W_\sigma(y|x)=\varphi_\sigma(y-x)$ and $\bar W_\sigma=\frac14\sum_{x=0}^3W_\sigma(\cdot|x)$.

\emph{Public dither.} Alice sends $X_j=(x_{\mathbf u,j}+D_j)\bmod4$, with a fresh dither $\mathbf D$ uniform on $\mathbb Z_4^n$ in every session, known to Bob and Eve. As $x\mapsto(x+d)\bmod4$ is not a PAM isometry, the dither changes the Euclidean code and in general Bob's error probability ($\{0,1\}$ becomes $\{3,0\}$ for $d=3$). Bob decodes with $W_{\sigma_{\mathrm B,j}}(y_j|(x_{\mathbf u,j}+D_j)\bmod4)$. The reliability results of~\cite{TITcomp} that we use are averages over the code ensemble and the dither for each state sequence independent of them, i.e., guarantees for a code drawn from a public seed, not for one deterministic code and all states~\cite[Rem.~2]{TITcomp}.

\emph{Stopping and feedback.} Symbols are sent in rounds ending at $n_1<n_2<\cdots$, $L_r=n_r-n_{r-1}$. A \emph{state-based} rule sets $\tau=\bar\tau(\mathsf S)$, e.g., the first round after which Bob's accumulated capacities exceed $(1+\delta)$ times the level sizes. A \emph{decoding-based} rule sets $\tau=\min(\tau_{\mathrm{dec}},\bar\tau)$, where $\tau_{\mathrm{dec}}$ is the first round at which Bob's verification test accepts; its public transcript $T$ (NACKs, ACK round, abort flag at the cap) takes $|\mathcal T|\le\bar r+1$ values, $\bar r$ being the number of rounds ending by $\bar\tau$, and must not depend on the hash seed. A $c$-bit CRC is an implementable test: it removes $c$ bits of source entropy and accepts a wrong word with probability about $2^{-c}$ per attempt.

\emph{Privacy amplification.} For secret-key agreement, Alice sends a uniform $\mathbf U$, and both parties compute $\mathbf M=f_{\mathbf h}(\mathbf U)$ after the session, with a $2$-universal family $f_{\mathbf h}:\mathbb F_2^k\to\mathbb F_2^{k_{\mathrm s}}$ and a fresh public seed $\mathbf h$, e.g., $f_{\mathbf h}(\mathbf u)=\mathbf u_{[1:k_{\mathrm s}]}\oplus\mathbf T_{\mathbf h}\mathbf u_{[k_{\mathrm s}+1:k]}$; $k_{\mathrm s}$ may depend on $\tau$ and on the published states. A secret \emph{message} of a length fixed in advance is sent by invert-then-encode~\cite{BTV2012}, drawing $\mathbf U$ uniformly from $f_{\mathbf h}^{-1}(\mathbf M)$. A length is the floor of the value given by the bound if that value is positive, and the key is empty otherwise.

\emph{Eve and the secrecy metric.} Eve observes $\mathbf Y_{\mathrm E}=(Y_{\mathrm E,j})_{j\le\tau}$, $T$, $\mathsf S$, $\mathbf h$, $\mathbf D$ and all code parameters, is computationally unbounded, and may combine all rounds. With $\mathbf V_{\mathrm E}=(\mathbf Y_{\mathrm E},T,\mathbf h,\mathbf D)$ and a uniform $\mathbf M$ independent of $(\mathbf h,\mathbf D,\mathsf S)$, $d_{\mathrm{TV}}=\frac12\|P_{\mathbf M\mathbf V_{\mathrm E}|\mathsf S}-P_{\mathrm{unif}}\times P_{\mathbf V_{\mathrm E}|\mathsf S}\|_1$. It averages over the public randomness: $d_{\mathrm{TV}}\le\varepsilon$ gives distance at most $\varepsilon/\alpha$ for all but a fraction $\alpha$ of seeds and dithers.

\section{Leakage Bounds Against Any Passive Adversary}\label{sec:bounds}
For $\theta\in(0,1]$ let
\begin{equation}\label{eq:Iren}
I_{1+\theta}(\sigma)=\frac1\theta\log\Big(\frac14\sum_{x=0}^{3}\int W_\sigma(y|x)^{1+\theta}\bar W_\sigma(y)^{-\theta}dy\Big),
\end{equation}
the order-$(1+\theta)$ R\'enyi information of uniform $4$-PAM, which tends to its mutual information $C_{\mathrm X}(\sigma)$ as $\theta\downarrow0$, and let $\mathcal L(\sigma)=\log\int\max_xW_\sigma(y|x)dy=\log(1+3\,\mathrm{erf}(\frac1{2\sqrt2\sigma}))$ be the maximal leakage~\cite{IWK2020}. Both vanish for $\sigma=\infty$; logarithms are to base $2$, and $A_n=\sum_{j\le n}I_{1+\theta}(\sigma_{\mathrm E,j})$.

\begin{theorem}[Rateless finite-length secrecy]\label{thm:secrecy}
Consider any code with $2^k$ equiprobable words $\mathbf u$ and a fixed stream $\mathbf x_{\mathbf u}\in\{0,1,2,3\}^\infty$ for each, privacy amplification with output length $k_{\mathrm s}$, and a state-based stopping rule. Fix $\mathsf S$.
\begin{enumerate}[label=(\alph*),leftmargin=*]
\item With the public dither, for every $\theta\in(0,1]$ and $\omega\ge0$,
\begin{equation}\label{eq:secrecyA}
d_{\mathrm{TV}}\le\tfrac12\,2^{-\frac12\left(k-k_{\mathrm s}-A_\tau-\omega\right)}+2\cdot2^{-\theta \omega}.
\end{equation}
\item Without dither, $d_{\mathrm{TV}}\le\frac12\,2^{-\frac12(k-k_{\mathrm s}-\sum_{j\le\tau}\mathcal{L}(\sigma_{\mathrm E,j}))}$.
\item With a decoding-based rule $\tau=\min(\tau_{\mathrm{dec}},\bar\tau)$ whose transcript $T$ is independent of $\mathbf h$ and takes at most $|\mathcal T|$ values, (a) and (b) hold with $\tau$ replaced by $\bar\tau$ and $k$ by $k-\log|\mathcal T|$.
\end{enumerate}
Moreover $I(\mathbf M;\mathbf V_{\mathrm E}|\mathsf S)\le d_{\mathrm{TV}}k_{\mathrm s}+(1+d_{\mathrm{TV}})h_{\mathrm b}(\frac{d_{\mathrm{TV}}}{1+d_{\mathrm{TV}}})$, and, if $\sigma_{\mathrm E,j}\ge\sigma_{\min}>0$, for every $\zeta>0$ there is $\theta>0$ such that the length $k_{\mathrm s}=k-\sum_{j\le\tau}C_{\mathrm X}(\sigma_{\mathrm E,j})-\zeta\tau-\theta^{-1}\log\frac4\varepsilon-2\log\frac2\varepsilon$ gives $d_{\mathrm{TV}}\le\varepsilon$.
\end{theorem}

\begin{IEEEproof}
Since $\mathbf h$ is independent of the source word and of everything Eve sees besides $\mathbf h$, it suffices to lower-bound the smooth average min-entropy of $\mathbf U$ given that view~\cite{DORS2008} and apply the leftover hash lemma: if a subnormalized measure obtained by removing a mass $\varepsilon_{\mathrm{sm}}$ has $\tilde H_\infty\ge\lambda$, the distance is at most $2\varepsilon_{\mathrm{sm}}+\frac12\sqrt{2^{k_{\mathrm s}-\lambda}}$~\cite{TSSR2011}.

(a) Here $\tau=n$ is fixed. Write the joint density of $(\mathbf U,\mathbf D,\mathbf Y_{\mathrm E})$ as $p(\mathbf u,\mathbf d,\mathbf y)=P_{\mathbf D}(\mathbf d)2^{-k}\prod_jW_{\sigma_{\mathrm E,j}}(y_j|(x_{\mathbf u,j}+d_j)\bmod4)$, let $\bar W^n=\prod_j\bar W_{\sigma_{\mathrm E,j}}$ and $\lambda=k-A_n-\omega$, and delete the set where $p>2^{-\lambda}P_{\mathbf D}(\mathbf d)\bar W^n(\mathbf y)$. On the rest, $\sum_{\mathbf d}\int\max_{\mathbf u}p'\le2^{-\lambda}$, i.e., $\tilde H_\infty(\mathbf U|\mathbf D,\mathbf Y_{\mathrm E})\ge\lambda$. For each fixed $\mathbf u$, averaging over $\mathbf D$ makes $\mathbf X$ i.i.d.\ uniform; hence the information densities $L_j=\log[W_{\sigma_{\mathrm E,j}}(Y_{\mathrm E,j}|X_j)/\bar W_{\sigma_{\mathrm E,j}}(Y_{\mathrm E,j})]$ are independent under the joint law with $\mathbb E[2^{\theta L_j}]=2^{\theta I_{1+\theta}(\sigma_{\mathrm E,j})}$, and Markov's inequality bounds the deleted mass by $2^{-\theta\omega}$.

(b) Without dither, $2^{-\tilde H_\infty(\mathbf U|\mathbf Y_{\mathrm E})}\le2^{-k}\int\max_{\mathbf x}\prod_jW_{\sigma_{\mathrm E,j}}(y_j|x_j)d\mathbf y=2^{-k}\prod_j2^{\mathcal L(\sigma_{\mathrm E,j})}$, because the maximum of a product of functions of distinct variables factorizes.

(c) Since the stream $\mathbf x_{\mathbf u}$ is fixed in advance, Eve's \emph{potential} observations $\mathbf Y_{\mathrm E}^{\bar\tau}$ up to the cap are well defined, and her actual view is a function of $(\mathbf Y_{\mathrm E}^{\bar\tau},\mathbf D,T,\mathbf h)$; discarding the unused tail cannot reduce secrecy. Apply the deletion of (a) once, at the full horizon $\bar\tau$, with $\lambda=k-A_{\bar\tau}-\omega$; the deleted mass is at most $2^{-\theta\omega}$. Append $T$ with its conditional law under the original experiment; then $p'(\mathbf u,\mathbf d,\mathbf y,t)\le p'(\mathbf u,\mathbf d,\mathbf y)$, and summing the maximum over $\mathbf u$ over the at most $|\mathcal T|$ values of $t$ gives $\tilde H_\infty(\mathbf U|\mathbf D,\mathbf Y^{\bar\tau}_{\mathrm E},T)\ge\lambda-\log|\mathcal T|$, whatever the dependence of $T$ on the message and on Bob's noise. As $T$ is independent of $\mathbf h$, the leftover hash lemma applies; no union bound over rounds is needed. The same argument gives the counterpart of (b).

The mutual-information bound is the continuity bound for conditional entropy~\cite[Lemma~2]{Winter2016}. For the secret length, $I_{1+\theta}$ is continuous on the compact set $[\sigma_{\min},\infty]$ and decreases to $C_{\mathrm X}$ as $\theta\downarrow0$; by Dini's theorem, $I_{1+\theta}\le C_{\mathrm X}+\zeta$ there for some $\theta>0$; the choice $\omega=\theta^{-1}\log\frac4\varepsilon$ makes each term of~\eqref{eq:secrecyA} at most $\varepsilon/2$.
\end{IEEEproof}

The bound is evaluated per session from the states; its exponent in $\theta$ is that of~\cite{Hayashi2011}. By~(c), at the same state-based cap a decoding-based acknowledgement costs $\log|\mathcal T|\le\log(\bar r+1)$ bits ($5$ bits for $31$ rounds); charging $A_{\bar\tau}$ instead of $A_\tau$ costs a further $A_{\bar\tau}-A_\tau$ bits, the price of not knowing $\tau$ in advance. The hypotheses matter: a rule that encoded a bit of $f_{\mathbf h}(\mathbf U)$ in the round index would reveal it whatever $k-k_{\mathrm s}$ is, and every further public message must be counted in $T$. Only the size and alphabet of the code and Eve's channel enter; knowledge of the lattice, dithers and hash, unbounded computation and joint decoding of all packets are covered. The dither makes the leakage rate approach $C_{\mathrm X}(\sigma_{\mathrm E})$: at $0$~dB, $\mathcal L=1.03$ bits, whereas $C_{\mathrm X}=0.49$ and $I_{1.05}=0.51$. The guarantee is finite-length $\varepsilon$-secrecy in total variation; asymptotic strong secrecy needs $\varepsilon_k\downarrow0$ with $\varepsilon_kk_{\mathrm s}\to0$, e.g., $\varepsilon_k=k^{-2}$, at a cost of $O(\log k)$ bits.

\section{Secret-Key Capacity of Degraded Sessions}\label{sec:cap}
Theorem~\ref{thm:secrecy}(a) holds verbatim for any alphabet on which the dithered symbols are uniform, in particular for the geometrically shaped $p$-ary rateless lattice codes of~\cite[Thm.~7]{TITcomp}, whose prefixes are points of a translate of a scaled rateless lattice and whose $p$ equiprobable points approximate a Gaussian. Let $I_g(\sigma)$ be the mutual information of such a constellation of power $P$. Since $2^{k_s}$ does not divide $p^{k_p}$, no balanced binary hash acts on all words; we use a fixed subset $\mathcal U_b$ of $2^b$ words, $b=\lfloor k_p\log p\rfloor$, labelled by binary strings, on which the full-code decoder errs at most twice as often on average, as $|\mathcal U_b|>p^{k_p}/2$. We assume bounds $\sigma_{\mathrm E,j}\ge\underline\sigma_{\mathrm E,j}$ published after the session, $\overline{\mathrm{SNR}}_{\mathrm E,j}=P/\underline\sigma_{\mathrm E,j}^2$ (a guard zone bounds path loss, not Rayleigh fading; Proposition~\ref{prop:adapt} treats statistical knowledge).

\begin{corollary}[Adaptive key generation]\label{cor:seccap}
In secret-key agreement with $\mathbf U$ uniform on $\mathcal U_b$, let Bob's received SNRs be at most $\mathrm{SNR}_{\max}$, let $\underline\sigma_{\mathrm E,j}$ lie in a compact subset of $(0,\infty]$, let Bob stop at the first round end $\tau$ with $\sum_{j\le\tau}I_g(\sigma_{\mathrm B,j})\ge(1+\delta)k_p\log p$, and let the length $L_\iota$ of the stopping round $\iota$ (with $\tau=n_\iota$) satisfy $L_\iota\le\ell_{k_p}$ with $\ell_{k_p}/k_p\to0$. For every $\epsilon>0$ there are $p$ and a constellation such that, for small $\delta$ and large $k_p$, Bob's ensemble-average error probability vanishes and a public key length with
\begin{equation}\label{eq:ach}
\frac{k_{\mathrm s}}\tau\ \ge\ \frac1\tau\sum_{j\le\tau}\tfrac12\log\frac{1+\mathrm{SNR}_{\mathrm B,j}}{1+\overline{\mathrm{SNR}}_{\mathrm E,j}}-\epsilon-\frac{L_\iota\log p+F_\varepsilon}{\tau}
\end{equation}
gives $d_{\mathrm{TV}}\le\varepsilon$, where $F_\varepsilon=\theta^{-1}\log\frac4\varepsilon+2\log\frac2\varepsilon+2$ includes integer rounding; the last term vanishes as $k_p\to\infty$.
\end{corollary}

\begin{IEEEproof}
Reliability is~\cite[Thm.~7]{TITcomp} at the state-based time $\tau$ (an ensemble guarantee for each fixed state sequence), restricted to $\mathcal U_b$. Let $B_n=\sum_{j\le n}I_g(\sigma_{\mathrm B,j})$. Minimality of $\tau$ gives $B_{n_{\iota-1}}<(1+\delta)k_p\log p$, and the stopping round adds at most $L_\iota\log p$, hence $k_p\log p>(B_\tau-L_\iota\log p)/(1+\delta)$; the round-length term cannot be bounded by $\log p$ unless rounds have one symbol. Theorem~\ref{thm:secrecy}(a) for this alphabet, with $I_{1+\theta}$ evaluated at $\underline\sigma_{\mathrm E,j}$ (it is nonincreasing in $\sigma$ by degradation) and $\theta$ chosen by Dini's theorem such that $I_{1+\theta}\le I_g+\zeta$ on the compact range, guarantees $k_{\mathrm s}=\lfloor b-\sum_{j\le\tau}I_g(\underline\sigma_{\mathrm E,j})-\zeta\tau-\theta^{-1}\log\frac4\varepsilon-2\log\frac2\varepsilon\rfloor\ge k_p\log p-\sum_{j\le\tau}I_g(\underline\sigma_{\mathrm E,j})-\zeta\tau-F_\varepsilon$ bits, using $b\ge k_p\log p-1$. Finally $I_g/(1+\delta)\ge I_g-\delta\log p$, $I_g(\sigma_{\mathrm B,j})\ge\frac12\log(1+\mathrm{SNR}_{\mathrm B,j})-\epsilon/3$ by~\cite[Thm.~7]{TITcomp}, $I_g(\underline\sigma_{\mathrm E,j})\le\frac12\log(1+\overline{\mathrm{SNR}}_{\mathrm E,j})$ by the Gaussian maximum-entropy bound; choosing $\zeta,\delta$ small gives~\eqref{eq:ach}. Since $I_g\le\log p$, $\tau\ge k_p$, and the last term of~\eqref{eq:ach} vanishes.
\end{IEEEproof}

Corollary~\ref{cor:seccap} concerns adaptive key generation; a predetermined secret message needs a length fixed in advance.

\begin{theorem}[Converse]\label{thm:converse}
Consider any scheme without a pre-shared secret key and without public discussion other than the publication of the states and of a state-based stopping notification, in which the public randomness $\mathsf Q$ (seeds, dithers, code choice) is independent of $(\mathbf M,\mathsf S)$, Alice's potential stream $(X_j)$ is independent of $\mathsf S$ given $\mathsf Q$, and $\mathbb E[X_j^2]\le P$. Fix $\mathsf S$ with $\sigma_{\mathrm E,j}\ge\sigma_{\mathrm B,j}$ for $j\le\tau$. If Bob's estimate $\hat{\mathbf M}(\mathbf Y_{\mathrm B},\mathsf Q)$ satisfies $\Pr\{\hat{\mathbf M}\neq\mathbf M\,|\,\mathsf S\}\le\varepsilon_{\mathrm B}$ and $I(\mathbf M;\mathbf Y_{\mathrm E},\mathsf Q\,|\,\mathsf S)\le\ell$ for a uniform $\mathbf M$, then
\begin{equation}\label{eq:conv}
(1-\varepsilon_{\mathrm B})\,k_{\mathrm s}\ \le\ \sum_{j\le\tau}\tfrac12\log\frac{1+\mathrm{SNR}_{\mathrm B,j}}{1+\mathrm{SNR}_{\mathrm E,j}}+\ell+1.
\end{equation}
With a pre-shared key of $\kappa$ bits, $\kappa$ is added to the right-hand side.
\end{theorem}

\begin{IEEEproof}
Condition on $\mathsf S$; then $n=\tau$ is fixed. By Fano's inequality, $(1-\varepsilon_{\mathrm B})k_{\mathrm s}\le I(\mathbf M;\mathbf Y_{\mathrm B}|\mathsf Q)+1$. Given $\mathsf Q$, reliability and secrecy depend only on the two marginal channels, because no transcript is computed from $\mathbf Y_{\mathrm B}$; we may therefore assume $Y_{\mathrm E,j}=Y_{\mathrm B,j}+Z''_j$ with independent $Z''_j\sim\mathcal N(0,\sigma_{\mathrm E,j}^2-\sigma_{\mathrm B,j}^2)$. The Markov chain $\mathbf M-\mathbf X-\mathbf Y_{\mathrm B}-\mathbf Y_{\mathrm E}$ given $\mathsf Q$ gives $I(\mathbf M;\mathbf Y_{\mathrm B}|\mathsf Q)\le I(\mathbf M;\mathbf Y_{\mathrm E}|\mathsf Q)+\sum_jI(X_j;Y_{\mathrm B,j}|Y_{\mathrm E,j},\mathsf Q)$, and $I(\mathbf M;\mathbf Y_{\mathrm E}|\mathsf Q)\le I(\mathbf M;\mathbf Y_{\mathrm E},\mathsf Q)\le\ell$ since $\mathbf M$ is independent of $\mathsf Q$. Given $\mathsf Q=\mathsf q$, the $j$-th term is at most $f_j(P_j(\mathsf q))$~\cite{LYCH1978}, where $P_j(\mathsf q)=\mathbb E[X_j^2|\mathsf Q=\mathsf q]$ and $f_j(v)=\frac12\log\frac{1+v/\sigma_{\mathrm B,j}^2}{1+v/\sigma_{\mathrm E,j}^2}$ is nondecreasing and concave for $\sigma_{\mathrm E,j}\ge\sigma_{\mathrm B,j}$. Jensen's inequality and $\mathbb E[P_j(\mathsf Q)]\le P$ give $\mathbb E[f_j(P_j(\mathsf Q))]\le f_j(P)$. A $\kappa$-bit key adds at most $\kappa$ to $I(\mathbf M;\mathbf Y_{\mathrm B}|\mathsf Q)-I(\mathbf M;\mathbf Y_{\mathrm E}|\mathsf Q)$.
\end{IEEEproof}

With tight bounds ($\overline{\mathrm{SNR}}_{\mathrm E,j}=\mathrm{SNR}_{\mathrm E,j}$) and short rounds, Corollary~\ref{cor:seccap} and Theorem~\ref{thm:converse} identify the accumulated Gaussian secrecy capacity of the realized session as the secret-key capacity of a \emph{degraded} session ($\sigma_{\mathrm E,j}\ge\sigma_{\mathrm B,j}$) under state-based stopping without further resources. The restrictions are essential (a pre-shared one-time pad gives positive rates with equal channels), and decoding-based stopping and non-degraded sessions remain open. Non-degraded rounds are not rare: with independent exponential SNRs whose means differ by the gap $G$, Eve is stronger in a round with probability $1/(1+G)$, $0.99\%$ for $20$~dB and $3.07\%$ for $15$~dB. For the $4$-PAM code below, $\sum_{j\le\tau}[C_{\mathrm X}(\sigma_{\mathrm B,j})-C_{\mathrm X}(\sigma_{\mathrm E,j})]$ is the uniform-input analogue.

\section{Statistical Eavesdropper Knowledge}\label{sec:adapt}
In practice Alice and Bob know only the law of Eve's states. Suppose that the states are constant within each round, $(\sigma_{\mathrm B,j},\sigma_{\mathrm E,j})=(\beta_r,\gamma_r)$ for $n_{r-1}<j\le n_r$, and that the pairs $(\beta_r,\gamma_r)$ are independent across rounds, with a specified conditional law of $\gamma_r$ given $\beta_r$; let $\boldsymbol\beta=(\beta_1,\beta_2,\dots)$.

\begin{proposition}[Key length with guaranteed outage]\label{prop:adapt}
In secret-key agreement with a state-based $\bar\tau=n_{\bar r}$, let $\Lambda_\theta=\sum_{r\le\bar r}L_rI_{1+\theta}(\gamma_r)$ and let $q_{1-\eta}(\boldsymbol\beta)$ be an upper $(1-\eta)$-quantile of its conditional law given $\boldsymbol\beta$. The key length $k_{\mathrm s}=\max\{0,\lfloor k-q_{1-\eta}(\boldsymbol\beta)-\theta^{-1}\log\frac4\varepsilon-2\log\frac2\varepsilon\rfloor\}$ satisfies $\Pr\{d_{\mathrm{TV}}>\varepsilon\,|\,\boldsymbol\beta\}\le\eta$ for every $\boldsymbol\beta$. Moreover,
\begin{equation}\label{eq:hoeff}
q_{1-\eta}(\boldsymbol\beta)\le\sum_{r\le\bar r}L_r\,\mathbb E[I_{1+\theta}(\gamma_r)|\beta_r]+\Big(2\ln\tfrac1\eta\sum_{r\le\bar r}L_r^2\Big)^{1/2}.
\end{equation}
\end{proposition}

\begin{IEEEproof}
Given $\boldsymbol\beta$, $\bar r$ is fixed, and Theorem~\ref{thm:secrecy}(a) with $\omega=\theta^{-1}\log\frac4\varepsilon$ gives $d_{\mathrm{TV}}\le\varepsilon$ whenever $\Lambda_\theta\le q_{1-\eta}(\boldsymbol\beta)$, an event of conditional probability at least $1-\eta$; an empty key has distance $0$. Given $\boldsymbol\beta$, the terms $L_rI_{1+\theta}(\gamma_r)$ are independent and lie in $[0,2L_r]$, because $W_\sigma(y|x)/\bar W_\sigma(y)\le4$; Hoeffding's inequality~\cite{Hoeffding1963} gives~\eqref{eq:hoeff}.
\end{IEEEproof}

This is a secrecy-outage guarantee, not secrecy in every session: if $d(\mathsf s)\le\varepsilon$ outside a set of states of probability $\eta$, the distance including the public states is $\mathbb E_{\mathsf S}d(\mathsf S)\le(1-\eta)\varepsilon+\eta$. With the long first round of Section~\ref{sec:num}, \eqref{eq:hoeff} is loose and we estimate the quantile by Monte Carlo; as a raw percentile may undershoot it, we use a tolerance quantile: with $m$ calibration samples, the $j$-th order statistic is an upper $(1-\eta)$-quantile with confidence $1-\alpha$ if $\Pr\{\mathrm{Bin}(m,1-\eta)\ge j\}\le\alpha$.

\begin{figure*}[t]
\centering
\includegraphics[width=\textwidth]{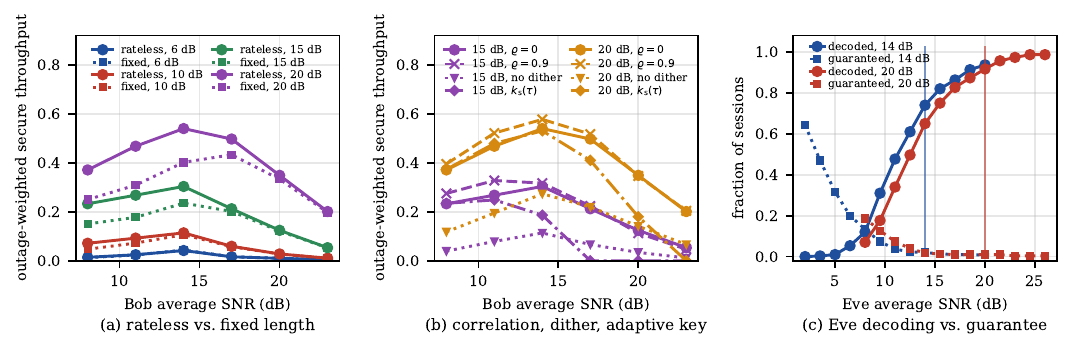}
\caption{Outage-weighted secure throughput over block Rayleigh fading, dithered transmission ($\varepsilon=10^{-6}$, $300$ sessions per point). (a) State-based rateless (solid) and best fixed-length (dotted) transmission with the same code and bound. (b) Correlation $\varrho$, no dither (Theorem~\ref{thm:secrecy}(b)), and key length $k_{\mathrm s}(\tau)$ for $5\%$ outage. (c) Sessions in which Eve's BP on all symbols recovers the lattice point, and in which Theorem~\ref{thm:secrecy}(a) guarantees $256$ bits; vertical lines: Bob's SNR.}
\label{fig:secrecy}
\end{figure*}

\section{What Spread Spectrum Can and Cannot Add}\label{sec:dsss}
Our earlier scheme~\cite{arxivPrev} modelled a DSSS layer as a fixed SNR penalty at Eve. The next proposition replaces that assumption.

\begin{proposition}[DSSS and secrecy]\label{prop:dsss}
Let each symbol $x_j$ be sent as $x_j\boldsymbol\psi_j$ over $\mathfrak s$ chips, $\boldsymbol\psi_j\in\{\pm1\}^{\mathfrak s}$, in white Gaussian chip noise of variances $\varsigma_{\mathrm B,j}^2$ and $\varsigma_{\mathrm E,j}^2$.
\begin{enumerate}[label=(\alph*),leftmargin=*]
\item If the sequences are known to Eve, Theorem~\ref{thm:secrecy} holds with $\sigma_{u,j}^2=\varsigma_{u,j}^2/\mathfrak s$, $u\in\{\mathrm B,\mathrm E\}$; DSSS leaves $\sigma_{\mathrm E,j}/\sigma_{\mathrm B,j}$ unchanged.
\item If the sequences are generated from a secret key $\mathbf K_{\mathrm{sp}}$ of $\kappa$ bits independent of $(\mathbf U,\mathbf h,\mathbf D)$, and $\mathbf V$ is Eve's chip-level view, then $\tilde H_\infty(\mathbf U|\mathbf V,\mathbf K_{\mathrm{sp}})\le\tilde H_\infty(\mathbf U|\mathbf V)\le\tilde H_\infty(\mathbf U|\mathbf V,\mathbf K_{\mathrm{sp}})+\kappa$, and $I(\mathbf U;\mathbf V,\mathbf K_{\mathrm{sp}})-I(\mathbf U;\mathbf V)=I(\mathbf U;\mathbf K_{\mathrm{sp}}|\mathbf V)\le\kappa$.
\end{enumerate}
\end{proposition}

\begin{IEEEproof}
(a) Projection on $\boldsymbol\psi_j/\sqrt{\mathfrak s}$ gives $\sqrt{\mathfrak s}\,x_j$ plus noise of variance $\varsigma_{u,j}^2$, a sufficient statistic. (b) Conditioning cannot increase average min-entropy, $\tilde H_\infty(\mathbf U|\mathbf V,\mathbf K)\ge\tilde H_\infty(\mathbf U|\mathbf V)-\log|\mathcal K|$~\cite[Lemma~2.2]{DORS2008}, and the identity is the chain rule.
\end{IEEEproof}

A keyed sequence reduces the joint leakage by at most its key length, a consumed resource; a per-symbol reduction of Eve's mutual information does not imply a per-symbol key requirement, since observations can describe the same source bits redundantly. In~(a) the symbol energy grows by the factor $\mathfrak s$ for both receivers; at fixed symbol energy there is no processing gain.

\section{Numerical Results}\label{sec:num}
\emph{Setup.} We use the Raptor lattice of~\cite{TITcomp} with $N=684$, $K=571$ (design~A, LT profile~M, \cite[Alg.~1]{TITcomp}), a first round of $684$ and further rounds of $114$ dimensions, at most $31$ rounds, and block Rayleigh fading with complex gains $h_{\mathrm E,r}=\varrho h_{\mathrm B,r}+\sqrt{1-\varrho^2}w_r$, $h_{\mathrm B,r},w_r\sim\mathcal{CN}(0,1)$ independent, $\mathrm{SNR}_{u,j}=|h_{u,r}|^2\overline{\mathrm{SNR}}_u$. Every session has a fresh public dither, and Bob and Eve decode the dithered symbols. Bob stops at a state-based time and uses multistage (MSD) BP with the level-capacity rule of margin $\delta$, or joint $\mathbb Z_4$-BP~\cite{TITcomp} with the rule $\sum_{j\le\tau}C_{\mathrm X}(\sigma_{\mathrm B,j})\ge(1+\delta)k$ (capacities by quadrature); decoding errors and cap failures count as losses. Eve's average SNR is $15$ or $20$~dB below Bob's (the gap); $\varrho=0$ unless stated. Each session is evaluated with Theorem~\ref{thm:secrecy}(a) at $\varepsilon=10^{-6}$, minimized over $12$ values of $\theta\in[0.01,1]$ and over $\omega$; if the bound exceeds $\varepsilon$ at the chosen $k_{\mathrm s}$, the session is a \emph{secrecy outage}. The \emph{outage-weighted} secure throughput is $\mathsf T_{\mathrm s}=k_{\mathrm s}\Pr\{\text{Bob correct, no secrecy outage}\}/\mathbb E[\tau]$, maximized over a fixed $k_{\mathrm s}$; it does not mean that every transmitted session is secure. Table~\ref{tab:sec} uses $1000$ sessions per point, with $\delta$ optimized over $\{0.25,0.35,0.45\}$ (MSD), $\{0.2,0.3,0.4\}$ (joint) and $0.3$ (ACK). In the ACK rows Bob decodes after every round up to $\bar\tau$ and stops at the first success, modelled as an ideal $16$-bit CRC: we charge $16$ bits ($k\to k-16$), use $|\mathcal T|=\bar r+1$, and subtract the undetected-error bound $(\bar r+1)2^{-16}<5\times10^{-4}$ from Bob's success probability. The fixed-length benchmark (same code and bound, optimized length and $k_{\mathrm s}$) declares Bob's success by the capacity criterion without decoding, which favours it. The ideal secure HARQ reference~\cite{TLSP2009} uses the same payload, rounds and $4$-PAM input, stops when $\sum_{j\le\tau}C_{\mathrm X}(\sigma_{\mathrm B,j})\ge k$ and declares secrecy when $\sum_{j\le\tau}C_{\mathrm X}(\sigma_{\mathrm E,j})\le k-k_{\mathrm s}$ (ideal decoding, asymptotic criterion). Bootstrap $95\%$ half-widths of $\mathsf T_{\mathrm s}$ are at most $0.028$~bits/dim.

\begin{table}[t]
\centering
\caption{Outage-weighted secure throughput (bits/dim) over block Rayleigh fading, dithered transmission, $\varrho=0$, $\varepsilon=10^{-6}$, $1000$ sessions per point}\label{tab:sec}
\footnotesize
\setlength{\tabcolsep}{2.5pt}
\begin{tabular}{llcccccc}
\hline
gap & Bob SNR (dB) & 8 & 11 & 14 & 17 & 20 & 23\\
\hline
\multirow{7}{*}{$20$~dB}
& ideal secure HARQ~\cite{TLSP2009} & 0.91 & 1.05 & 1.09 & 1.02 & 0.86 & 0.65\\
& joint BP, MI criterion & 0.64 & 0.77 & 0.84 & 0.80 & 0.67 & 0.52\\
& joint BP, Thm.~1(a) & 0.56 & 0.65 & 0.70 & 0.64 & 0.52 & 0.38\\
& joint BP, CRC-ACK, Thm.~1(c) & 0.61 & 0.70 & 0.75 & 0.68 & 0.51 & 0.32\\
& MSD BP, Thm.~1(a) & 0.36 & 0.50 & 0.57 & 0.54 & 0.41 & 0.26\\
& fixed, joint criterion, Thm.~1(a) & 0.42 & 0.47 & 0.54 & 0.52 & 0.45 & 0.36\\
& fixed, MSD criterion, Thm.~1(a) & 0.25 & 0.35 & 0.43 & 0.46 & 0.39 & 0.25\\
\hline
\multirow{7}{*}{$15$~dB}
& ideal secure HARQ~\cite{TLSP2009} & 0.75 & 0.80 & 0.76 & 0.64 & 0.48 & 0.32\\
& joint BP, MI criterion & 0.51 & 0.56 & 0.57 & 0.48 & 0.34 & 0.23\\
& joint BP, Thm.~1(a) & 0.41 & 0.44 & 0.43 & 0.35 & 0.24 & 0.15\\
& joint BP, CRC-ACK, Thm.~1(c) & 0.44 & 0.47 & 0.44 & 0.32 & 0.20 & 0.10\\
& MSD BP, Thm.~1(a) & 0.23 & 0.31 & 0.33 & 0.26 & 0.15 & 0.08\\
& fixed, joint criterion, Thm.~1(a) & 0.29 & 0.31 & 0.32 & 0.27 & 0.22 & 0.15\\
& fixed, MSD criterion, Thm.~1(a) & 0.16 & 0.20 & 0.25 & 0.24 & 0.15 & 0.08\\
\hline
\end{tabular}
\end{table}

\emph{Coding alone is not secrecy.} In all $8089$ simulated sessions in which Eve's BP recovered the transmitted lattice point, Theorem~\ref{thm:secrecy} guaranteed no secret bit. With a $6$~dB gap at Bob's $14$~dB, Eve recovers the lattice point in $12\%$ of the sessions (Fig.~\ref{fig:secrecy}(c)).

\emph{Rateless versus fixed length.} With MSD BP, $\delta=0.35$ and a $20$~dB gap, the rateless scheme achieves $0.37$, $0.47$, $0.54$ and $0.50$ bits/dim at Bob SNRs of $8$, $11$, $14$ and $17$~dB, and the best fixed-length transmission $0.25$, $0.31$, $0.40$ and $0.43$ (Fig.~\ref{fig:secrecy}(a)). The gain of $15$--$51\%$ is the leakage adaptivity: Alice stops early when Bob's channel is good. At high SNR both receivers approach the $2$-bit limit of $4$-PAM; shaped codes (Corollary~\ref{cor:seccap}) would remove it, but lack a practical decoder.

\emph{Joint decoding, acknowledgements and the gap to the ideal.} Joint BP at Bob raises the throughput over MSD BP by $18$--$55\%$ for a $20$~dB gap and $30$--$90\%$ for a $15$~dB gap (Table~\ref{tab:sec}), because it shortens $\mathbb E[\tau]$ (from $2335$ to $1565$ dimensions at $8$~dB), and every dimension saved is withheld from Eve. With joint BP the rateless scheme beats the fixed-length scheme with the same criterion by $4$--$43\%$. Relative to the ideal, joint BP with state-based stopping keeps $68$--$80\%$ (MI criterion), and the finite-length bound a further $65$--$87\%$. The CRC-based acknowledgement with the bound of Theorem~\ref{thm:secrecy}(c) shortens $\mathbb E[\tau]$ further (from $987$ to $911$ dimensions at $14$~dB) and raises the throughput by $6$--$9\%$ up to $17$~dB for a $20$~dB gap; at higher SNR its costs ($\log|\mathcal T|$, CRC, leakage up to the cap) dominate. For a $20$~dB gap and Bob SNRs of $8$--$20$~dB, the better of the two stopping rules reaches $60$--$69\%$ of the ideal benchmark ($66$--$69\%$ with the acknowledgement up to $17$~dB), and $58\%$ at $23$~dB; for a $15$~dB gap, $54$--$59\%$ up to $17$~dB. At the optimum, the secrecy-outage probability of the joint Thm.~1(a) row grows with the SNR, from $4\%$ to $33\%$ ($20$~dB gap) and from $10\%$ to $66\%$ ($15$~dB), since a larger $k_{\mathrm s}$ is then worth more outages.

\emph{Dither, correlation and adaptive keys.} The public dither doubles the throughput compared with undithered transmission evaluated with Theorem~\ref{thm:secrecy}(b): $0.54$ versus $0.27$ bits/dim at $14$~dB with a $20$~dB gap (Fig.~\ref{fig:secrecy}(b)). Strong Bob--Eve correlation slightly \emph{increases} it at moderate SNR ($0.52$ versus $0.47$ at $11$~dB), because sessions in which Bob stops early are then also those in which Eve is strong. For Proposition~\ref{prop:adapt} with $\eta=5\%$ and $\varrho=0$ (the law of $\Lambda_\theta$ given $\boldsymbol\beta$ then depends only on $\bar r$) we use $m=4000$ calibration samples per stopping round and $j=3840$, i.e., confidence $95\%$ simultaneously over the $31$ possible rounds; on $300$ independent validation sessions the scheme keeps $0.37$--$0.53$ bits/dim for a $20$~dB gap between $8$ and $14$~dB, with measured outage $2.0$--$6.3\%$ between $8$ and $20$~dB, compatible with the $5\%$ target (standard error $1.3$ points). Gaps of $10$~dB or less leave little provable secrecy under independent Rayleigh fading~\cite{BBRM2008}.

\section{Conclusion}
Rateless lattice codes with a public dither and privacy amplification give per-state, finite-length secrecy against the strongest passive adversary, also with acknowledgements on decoding success, at a timing cost of the logarithm of the number of feedback outcomes. For degraded sessions with state-based stopping, adaptive key generation with shaped rateless lattice codes attains the accumulated Gaussian secrecy capacity, which our converse shows to be the limit. In simulations, joint decoding at Bob matters for secrecy as much as for reliability. Converses for non-degraded sessions and decoding-based stopping, and active adversaries, are open.

\bibliographystyle{IEEEtran}
\bibliography{refs}
\end{document}